\documentclass [twocolumn,epsfig,pre,10pt,floatfix,amsmath,amssym]{revtex4}
\usepackage[dvips]{epsfig}

\usepackage{float}

\begin{document}

\title{Self-Consistent Field Theory of Multiply-Branched Block Copolymer Melts}
\author{Gregory M. Grason}
\author{Randall D. Kamien}
\affiliation{Department of Physics and Astronomy, University of Pennsylvania, Philadelphia, PA 19104-6396, USA}

\begin{abstract}
We present a numerical algorithm to evaluate the self-consistent field theory for melts composed of block copolymers with multiply-branched architecture.  We present results for the case of branched copolymers with doubly-functional groups for multiple branching generations.  We discuss the stability of the cubic phase of spherical micelles, the A15 phase, as a consequence of tendency of the AB interfaces to conform to the polyhedral environment of the Voronoi cell of the micelle lattice.
\end{abstract}
\date{\today}
\maketitle

\section{Introduction}
Block copolymers provide an ideal route to engineering well-controlled structure on nanometer length scales \cite{bates_sci_91, bates_fred_arpc_90}.  Through control over the chemical architecture, these systems can be tuned to self-assemble into periodic structures of an astounding variety.  A plethora of new phases and structures have been identified in dilute diblock systems \cite{bates_sci_03,discher_polymer}, triblock systems \cite{triblock}, and confined diblocks \cite{russell}.  One might think there is hardly more to say about melts of the simplest of block copolymer architectures, the neat linear AB diblock copolymer.  It is well known that these linear diblock copolymers display a host of ordered phases:  spheres, cylinders, lamella and the bicontinuous gyroid \cite{matsen_jphys_02}.   However, we have argued \cite{grason_didonna_kamien_prl_03,grason_kamien_macro_04}  that the tendency to minimize the AB interfacial area should stabilize a new cubic phase with $Pm\overline{3}n$ symmetry, the A15 lattice.  The subsequent synthesis and characterization of PEO-docosyl dendrimeric diblocks  corroborated our prediction \cite{cho_science_04} and was in agreement with the self-consistent field theory (SCFT) phase diagram for miktoarm star copolymers \cite{grason_kamien_macro_04}.   In this article, we provide the details of SCFT for branched architectures and, to our knowledge, the first SCFT phase diagrams for 
true, multiply-branched dendritic diblock copolymers.  

The serial development of new chemical synthesis routes is typically a costly and slow means for exploring the consequences of novel copolymer architectures.  It is therefore desirable to develop theoretical tools for the efficient computation the phase behavior which can systematically map out novel phase properties for a broad class copolymer architectures.  
Milner and Olmsted developed a strong-segregation theory (SST) approach to the phase behavior of A$_n$B$_m$ miktoarm star copolymer melts, applicable in the $\chi N \rightarrow \infty$ limit, where $\chi$ is the Flory-Huggins immiscibility parameter and $N$ is the total number of chemical segments in the copolymer \cite{milner_macro_94, olmsted_milner_macro_98}.  For asymmetric copolymers, say for $n>m$, the effective spring constant of the more abundant polymer block is $n^2/m^2$ times more stiff than other block.   Such asymmetry leads an enhanced stability of phases with strong interfacial curvature, and thus, spherical and cylindrical micelles are predicted to dominate the phase behavior for large molecular asymmetry \cite{milner_macro_94}.  Fredrickson and Frischknecht introduced an approximate SST approach to multiply-branched dendritic copolymers \cite{fred_macro_99}, and Pickett developed a more refined self-consistent brush analysis for dendritic copolymer melts \cite{pickett_macro_02}.  Both works showed a similar increase in stability of high interface curvature phases.  Despite the analytic transparency of these SST calculations, the results of these calculations are predicated on many assumptions about the detailed structure of the micellar aggregates.  In particular, certain assumptions must be made concerning the interfacial shape and the direction in which copolymer chains stretch in the aggregates \cite{grason_kamien_macro_04, olmsted_milner_macro_98}.  Because the free energy differences between phases are small, the presence of these undetermined degrees of freedom makes the task of locating the true free energy ground state analytically cumbersome, if not impossible.

\begin{figure}
\epsfig{file=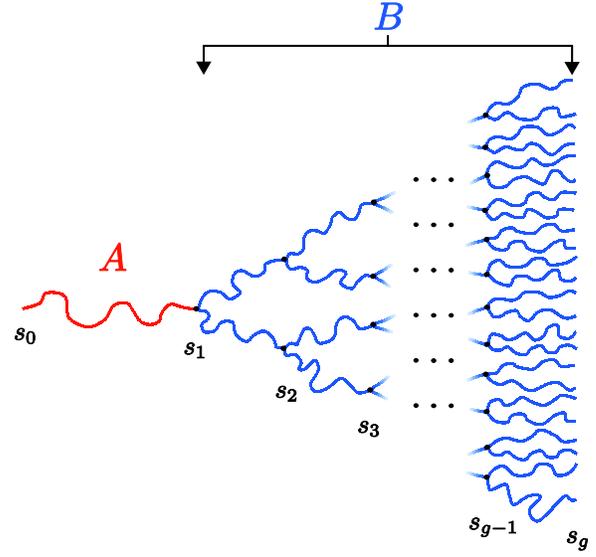,width=3.0in} \center \caption{A
schematic of the branched molecular architecture.  The first
generation A-block contains $f N$ segments.  Each higher
generation B-block is composed of $(1-f) N /
\mathcal{N}_{\mathrm{B}}$ segments. Here, the branching of each
generation is 2.  In the mean-field approximation, it is necessary to define only a single coordinate,
$s_\alpha$, for the set of branching points of the
$\alpha^{\rm{th}}$ generation.} \label{fig: branched}
\end{figure}

In Section II we present the theoretical derivation of the SCFT for multiply-branched copolymer melts from the full classical partition function of this system.  We present an algorithm for the SCFT of block copolymers within a specific class of multiply-branched architectures (see Figure \ref{fig: branched}).  Like the SCFT approaches of Matsen and Schick for linear AB diblock copolymers \cite{matsen_schick_prl_94} and for (AB)$_n$ starblock copolymers \cite{matsen_schick_macro_94}, this approach makes no approximation beyond the approximation of mean-field in the monomer concentration profile.  Therefore, this SCFT fully captures the copolymer chain fluctuations in the presence of the average concentration profile of constituent monomers.  Moreover, this approach efficiently minimizes over all possible copolymer configurations, precluding the variational assumptions often necessary in the SST calculations.  Finally, a numerical implementation of SCFT is not limited to the infinite $\chi N$ parameter range.  Given an arbitrary amount of computing time the equilibrium phase can be determined for any finite value of $\chi N$.  Practically, SCFT provides an efficient means of computing the mean field free energy of most phases for $\chi N \lesssim 100$ \cite{matsen_bates_macro_96_1}.
In Section III we present the results of an application of SCFT to a series of branched copolymer melts within a specific class of this structural motif, specifically, copolymers which branch doubly with each successive generation.  We discuss these results in context of elastically asymmetric copolymer melts and the geometry of the AB interface.  We conclude in Section IV.

\section{Self-Consistent Field Theory for Branched Architectures}

Our approach to multiply-branched diblocks is an extension of the SCFT
approach to linear diblocks and starblock copolymers pioneered by
Matsen and Schick \cite{matsen_schick_prl_94,matsen_schick_macro_94}.  While the derivation of the mean-field
free energy for the mulitply-branched system follows directly from the
results for the linear and starblock architectures, we present its
full derivation here since the subsequent evaluation of that free
energy requires a slightly more generalized approach.
Nevertheless, where possible, we attempt to keep the notation
consistent with theirs.

We consider a system of total volume, $V$, containing $n$ branched
copolymers. These copolymers are each composed of $N$ total
segments.  Without loss of generality, we define the segment volume for both monomer type to
be $\rho_0^{-1}$, so that the total volume of the system is $V =
\frac{n N}{\rho_0}$.  The statistical segment lengths for the A
and B-type monomers are denoted by $a_{\mathrm{A}}$ and
$a_{\mathrm{B}}$.  The volume fraction of A-type monomer in the
system is denoted by $f$. Thus, each chain is composed of $f N$
A-type segments and $(1-f) N$ B-type segments.
The architecture of our molecule is shown in Figure \ref{fig:
branched}.  The first generation is a single A-block.  Grafted
onto this are $(g-1)$ generations of equal length B-blocks.  The
branching of the $\alpha^{\mathrm{th}}$ generation is given by
$\eta_{\alpha}$ so that the total number of $B$ blocks,
$\mathcal{N}_{\mathrm{B}}$, is given by,
\begin{equation}
\mathcal{N}_{\mathrm{B}} = \eta_{2}( 1 +\eta_{3}(1+\eta_{4}(\ldots
(1+\eta_{g-1}(1+\eta_{g})) \ldots ) \ .
\end{equation}
We define a coordinate along the polymer, $s$, so that within any
chain portion of length $\Delta s$ there are $(\Delta s )N$
segments. Thus, in these coordinates, the length of the A block is
given by $\Delta s_{\mathrm{A}}=f$ and that of the B sections is
given by $\Delta s_{\mathrm{B}}=(1-f) / \mathcal{N}_{\mathrm{B}}$.

A particular melt configuration is specified by $n$ branched
curves in space, $\textbf{r}_\beta(s)$, 
the course-grained position of the $(sN) ^ {\mathrm{th}}$ segment
of the $\beta^{\mathrm{th}}$ chain.  At this point, we do not
introduce an explicit parameterization of the full branched
configuration.  It suffices to demand that the first generation
curve is joined to $\eta_2$ second generation curves which are
each joined to $\eta_3$ third curves, { \it etc.}.  Given this set of
branched curves, we define the dimensionless segment density
operators,
\begin{equation}
\label{eq: pA}
\hat{\phi}_{\mathrm{A}}(\textbf{r}) \equiv
\frac{N}{\rho_0} \sum_{\beta=1}^{n} \int_0^1 \!\!\! ds \ \gamma(s)
\ \delta(\textbf{r}-\textbf{r}_\beta(s))  \ ,
\end{equation}
\begin{equation}
\label{eq: pB}
\hat{\phi}_{\mathrm{B}}(\textbf{r}) \equiv
\frac{N}{\rho_0} \sum_{\beta=1}^{n} \int_0^1 \!\!\! ds \
(1-\gamma(s)) \ \delta(\textbf{r}-\textbf{r}_\beta(s))  \ ,
\end{equation}
where $\gamma(s)$ is a function which is equal to 1 when $s$ lies
along an A portion of the chain and 0 when $s$ is along a B portion
of the chain, and the integration range is over the entire
branched curve.  In a neat system, the allowed melt configurations are incompressible, and thus
we are constrained to consider configurations for which
\begin{equation}
\hat{\phi}_{\mathrm{A}}(\textbf{r})+\hat{\phi}_{\mathrm{B}}(\textbf{r})
= 1 \ .
\end{equation}

The full partition function for the melt is the
functional integral over $n$ branched curves:
\begin{multline}
\label{eq: z1} \mathcal{Z} = \frac{1}{n!} \int \prod_{\beta=1}^n
[d\textbf{r}_\beta] \
\delta[1-\hat{\phi}_{\mathrm{A}}(\textbf{r})-\hat{\phi}_{\mathrm{B}}(\textbf{r})] \\
\times \ \exp \Bigg\{ - \frac{3}{2Na^2}\int_{0}^{1} ds \
\bigg[\gamma(s) + \kappa^2(1-\gamma(s)) \bigg]
|\dot{\textbf{r}}_\beta(s)|^2 \\
 - \chi \rho_0 \int d\textbf{r} \
\hat{\phi}_{\mathrm{A}}(\textbf{r})
\hat{\phi}_{\mathrm{B}}(\textbf{r}) \Bigg\} \ ,
\end{multline}
where a normalization factor is absorbed into the functional
measure, $[d\textbf{r}_\beta]$,
$\dot{\textbf{r}}(s)=d\textbf{r}(s)/ds$,  $\kappa \equiv a_{\mathrm{A}}/a_{\mathrm{B}}$ measures the relative length of the A and B segments, and $a \equiv a_{\mathrm{A}}$.  The Flory-Huggins parameter,
$\chi$, characterizes the repulsive interaction between unlike
monomers.

We can use the identity $ \int [d \Phi_{\mathrm{A,B}}] \ \delta
[\Phi_{\mathrm{A,B}}(\textbf{r}) -
\hat{\phi}_{\mathrm{A,B}}(\textbf{r})] = 1$ to transform (\ref{eq:
z1}) into a functional integral over the monomer distributions.
Introducing fields conjugate to the total and individual segment
concentrations, we have explicit representations of the delta-functionals,
\begin{multline}
\label{eq: dphitot}
\delta[1-\hat{\phi}_{\mathrm{A}}(\textbf{r})-\hat{\phi}_{\mathrm{B}}(\textbf{r})]
= \\
\int [d \Xi] \ \exp \Bigg\{ \frac{\rho_0}{N} \int d\textbf{r} \
\Xi(\textbf{r}) [ 1 -
\hat{\phi}_{\mathrm{A}}(\textbf{r})-\hat{\phi}_{\mathrm{B}}(\textbf{r})
] \Bigg\} \ ,
\end{multline}
and,
\begin{multline}
\label{eq: dphiAB}
\delta[\Phi_{\mathrm{A,B}}(\textbf{r})-\hat{\phi}_{\mathrm{A,B}}(\textbf{r})]
= \\
\int [d W_{\mathrm{A,B}}] \ \exp \Bigg\{ \frac{\rho_0}{N} \int
d\textbf{r} \ W_{\mathrm{A,B}}(\textbf{r})
[\Phi_{\mathrm{A,B}}(\textbf{r})-\hat{\phi}_{\mathrm{A,B}}(\textbf{r})]
\Bigg\} \ ,
\end{multline}
where the limits of integration of the conjugate fields are $\pm i
\infty $.  Inserting these representations and the above identity
into (\ref{eq: z1}) and integrating over the delta functions in
(\ref{eq: pA}) and (\ref{eq: pB}), the full partition function is
given by,
\begin{multline}
\label{eq: z2}
 \mathcal{Z} =  \frac{1}{n!} \int [d \Xi][d W_{\mathrm{A}}][d
W_{\mathrm{B}}][d \Phi_{\mathrm{A}}][d \Phi_{\mathrm{B}}] \\
\times
(\mathcal{Q}[W_{\mathrm{A}}(\textbf{r}),W_{\mathrm{B}}(\textbf{r})])^n
 \exp \Bigg\{-\frac{n}{V} \int d\textbf{r} \ \bigg[ \chi N
\Phi_{\mathrm{A}}(\textbf{r}) \Phi_{\mathrm{B}}(\textbf{r}) \\ -
W_{\mathrm{A}}(\textbf{r}) \Phi_{\mathrm{A}}(\textbf{r})  -
W_{\mathrm{B}}(\textbf{r}) \Phi_{\mathrm{B}}(\textbf{r}) -
\Xi(\textbf{r}) [ 1 - \Phi_{\mathrm{A}}(\textbf{r}) -
\Phi_{\mathrm{B}}(\textbf{r})] \bigg] \Bigg\} \ ,
\end{multline}
where
$\mathcal{Q}[W_{\mathrm{A}}(\textbf{r}),W_{\mathrm{B}}(\textbf{r})]$
is the partition function for a single non-interacting, branched
chain subject to the spatial field, $W_{\mathrm{A}}(\textbf{r})$
acting on first generation of the chain and
$W_{\mathrm{B}}(\textbf{r})$ acting on the higher generations:
\begin{multline}
\label{eq: Q}
\mathcal{Q}[W_{\mathrm{A}}(\textbf{r}),W_{\mathrm{B}}(\textbf{r})]
= \int \prod_{\beta=1}^n[d\textbf{r}_\beta] \\
\times \ \exp \Bigg\{ - \int_{0}^{1} ds \ \bigg[\gamma(s) \bigg(
\frac{3}{2Na^2}|\dot{\textbf{r}}_\beta(s)|^2 +
W_{\mathrm{A}}(\textbf{r}_{\beta}(s)) \bigg) \\
+ (1-\gamma(s)) \bigg(
\frac{3\kappa^2}{2Na^2}|\dot{\textbf{r}}_\beta(s)|^2 +
W_{\mathrm{B}}(\textbf{r}_{\beta}(s)) \bigg) \bigg] \Bigg\} \ .
\end{multline}
In general, it is not possible to evaluate the functional
integrals in (\ref{eq: z2}).  Nevertheless, in the limit where $N$
is large, fluctuation contributions to the partition function are
small, and the integral is dominated by its saddle-point, where
the free energy per chain, $- \frac{k_B T}{n} \ln\mathcal{Z}$, is
minimal \cite{helfand_fred_jcp_87, fred_macro_review_02} .  The saddle-point approximation, of course, yields the mean-field results.

To obtain the mean-field result, we solve for the field
configurations, $[\phi_{\mathrm{A}}(\textbf{r})$,
$\phi_{\mathrm{B}}(\textbf{r})$, $w_{\mathrm{A}}(\textbf{r})$,
$w_{\mathrm{B}}(\textbf{r})$, $\xi(\textbf{r})]$,which minimize
the free energy (that is, the {\it lower-case} functions are the extremal
values of the {\it upper-case} functions).  Minimizing with respect to
$\Phi_{\mathrm{A}}(\textbf{r})$, $\Phi_{\mathrm{B}}(\textbf{r})$
and $\Xi(\textbf{r})$, respectively, we obtain the
mean-field equations:
\begin{equation}
\label{eq: wA} w_{\mathrm{A}}(\textbf{r})=\chi N
\phi_{\mathrm{B}}(\textbf{r}) + \xi(\textbf{r}) \ ,
\end{equation}
\begin{equation}
\label{eq: wB} w_{\mathrm{B}}(\textbf{r})=\chi N
\phi_{\mathrm{A}}(\textbf{r}) + \xi(\textbf{r}) \ ,
\end{equation}
\begin{equation}
\label{eq: incomp} 1=\phi_{\mathrm{A}}(\textbf{r}) +
\phi_{\mathrm{B}}(\textbf{r}) \ .
\end{equation}
Minimizing with respect to $W_{\mathrm{A}}(\textbf{r})$ and
$W_{\mathrm{B}}(\textbf{r})$, respectively, we find expressions
for the mean-field densities,
\begin{equation}
\label{eq: pAB} \phi_{\mathrm{A,B}}(\textbf{r}) = -
\frac{nN}{\rho_{0} \mathcal{Q}} \frac{\delta \mathcal{Q}}{\delta
w_{A,B}(\textbf{r})} \ ,
\end{equation}
where we have defined $\mathcal{Q} \equiv
\mathcal{Q}[w_{\mathrm{A}}(\textbf{r}),w_{\mathrm{B}}(\textbf{r})]$.

Upon inspection, it is clear how these relations constitute the
mean-field theory result of the full problem.  We have replaced
the problem of multiply-branched chains mutually interacting, with the problem
of non-interacting chains subject to the fields
$w_{\mathrm{A}}(\textbf{r})$ and $w_{\mathrm{B}}(\textbf{r})$.
These fields are chosen to represent the mean-field interactions
produced by the monomer distributions
$\phi_{\mathrm{A}}(\textbf{r})$ and
$\phi_{\mathrm{B}}(\textbf{r})$,  That is, from (\ref{eq: wA}) and
(\ref{eq: wB}) it is clear that A-type (B-type) monomers
experience a repulsion proportional to $\chi N$ times the local
density of B-type (A-type) monomers and a repulsion due to the
overall incompressibility of the system, given by
$\xi(\textbf{r})$.  Because the mean-field incompressibility
constraint, (\ref{eq: incomp}), depends only on the total
monomer density, $\xi(\textbf{r})$ contributes equally to both
potentials, $w_{\mathrm{A}}(\textbf{r})$ and
$w_{\mathrm{B}}(\textbf{r})$.  Hence, we see that
$\xi(\textbf{r})$ is simply the Lagrange-multiplier field which
allows us to fix the combined, local segment concentration to
$\rho_0$.  Moreover, the average segment distributions, (\ref{eq:
pAB}), are simply the average distributions produced by
non-interacting chains subject to the fields
$w_{\mathrm{A}}(\textbf{r})$ and $w_{\mathrm{B}}(\textbf{r})$.
Thus, eqns. (\ref{eq: wA})-(\ref{eq: pAB}) provide a fully
self-consistent set of equations, which can be solved to yield the
mean-field result. Once the $w_{\mathrm{A}}(\textbf{r})$ and
$w_{\mathrm{B}}(\textbf{r})$ are found, we can compute the mean-field free energy per
chain,
\begin{multline}
\label{eq: Freal} \frac{F}{n k_B T} = - \ln\mathcal{Q} - V^{-1}
\int d\textbf{r} [ w_{\mathrm{A}}(\textbf{r}) \phi_A(\textbf{r}) +
w_{\mathrm{B}}(\textbf{r}) \phi_B(\textbf{r})] \\
+ V^{-1} \int d\textbf{r} \ \chi N \phi_A(\textbf{r})
\phi_B(\textbf{r}) \ .
\end{multline}
The first line of (\ref{eq: Freal}) gives the entropy per branched
chain, and the second line gives the enthalpic, or interaction,
contribution to the free energy.

For a given set of monomer potentials,
$w_{\mathrm{A}}(\textbf{r})$ and $w_{\mathrm{B}}(\textbf{r})$,
$\mathcal{Q}$ can be evaluated.  We start by defining the
Green's function, or propagator, for a continuous, unbranched
portion of the chain,
\begin{multline}
\label{eq: G} G(\textbf{r}_i, s_i ;
\textbf{r}_f, s_f) \equiv \int_{\textbf{r}_i}^{\textbf{r}_f} [d
\textbf{r}_\beta] \\
\times \ \exp \bigg\{ - \int_{s_i}^{s_f} ds \bigg[ \frac{3}{2Na^2}
|\dot{\textbf{r}}_\beta(s)|^2 + w_A(\textbf{r}_\beta(s)) \bigg]
\gamma(s) \\
+ \bigg[ \frac{3 \kappa^2}{2Na^2} |\dot{\textbf{r}}_\beta(s)|^2 +
w_B(\textbf{r}_\beta(s)) \bigg] \big(1-\gamma(s) \big) \bigg\} \ ,
\end{multline}
where this path integral is carried out over all paths,
$\textbf{r}_\beta(s)$, which such that
$\textbf{r}_\beta(s_i)=\textbf{r}_i$ and
$\textbf{r}_\beta(s_f)=\textbf{r}_f$.  We absorb a normalization
into $[d \textbf{r}_\beta]$ so that the integral of the propagator
over the coordinates $\textbf{r}_i$ and $\textbf{r}_f$ is
independent of arc length, $s_f-s_i$.  This is the same as
demanding that the probability of any portion of this chain having
\emph{any} configuration (in the absence of external potentials)
is independent of the number of segments it contains.  Note that
$G (\textbf{r}_i, s_i ; \textbf{r}_f, s_f)$ is
identical to the imaginary-time quantum mechanical amplitude (with $s \rightarrow
-it)$ for a particle of mass, $N a^2 /3$ (or
$N a^2 /3 \kappa^2$ when $\gamma(s)=0$), in the potential,
$-w_A(\textbf{r})$ (or $-w_B(\textbf{r})$ when $\gamma(s)=0$)
moving from $\textbf{r}_i$ at the initial ``time," $s_i$, to
$\textbf{r}_i$ at a later ``time," $s_f$.  Therefore, we know that
$G (\textbf{r}_i, s_i ; \textbf{r}_f, s_f)$
obeys the imaginary-time Schr\"{o}dinger equation, or diffusion
equation and, unlike its interpretation in quantum mechanics, represents a
probability and \emph{not} an amplitude.  We make explicit use of this fact below.

\begin{figure}
\epsfig{file=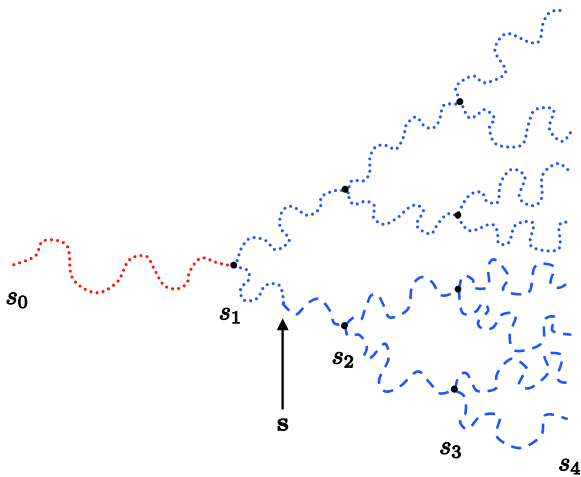,width=3.0in} \center \caption{A
schematic representation of the probability captured by the
end-distribution functions, $q^\dag(\textbf{r},s)$ and
$q(\textbf{r},s)$, for a 4-generation molecule.  For the point,
$s$, $q^\dag(\textbf{r},s)$ is proportional to the probability
that the dashed portion of the chain has diffused to the position,
$\textbf{r}$.  For the same point, $q(\textbf{r},s)$ is
proportional to the probability that the dotted portion of the
chain has diffused to the the same position. The probability that the point is at
{\bf r} at $s$ is the product of $q$ and $q^\dag$.} \label{fig: qs}
\end{figure}

To capture the branched architecture of the chain we define the
end-distribution functions.  These functions compute the
statistical weight of a chain diffusing along its trajectory to
some position in space.  That is, we define a function,
$q^\dag(\textbf{r},s)$, which is proportional to the probability
that the branched chain diffused from one of its free ends at
$s_g$, where $s_\alpha$ is the length coordinate corresponding to
the branching point of $\alpha^{\rm{th}}$ generation (see
Fig.~\ref{fig: branched}). Note that for $s_{g-1} < s < s_{g}$
this function is simply the probability that an unbranched chain
diffused from its free end to $s$ at some position, $\textbf{r}$.
But if $s_{g-1} < s < s_{g-2}$, then $q^\dag(\textbf{r},s)$ is
proportional to the probability that $\eta_g$ free ends diffused
from $s_g$ to some intermediate position, say $\textbf{r}_{g-1}$,
at $s_{g-1}$ and then diffused on to $\textbf{r}$ at $s$ (see Fig.
~\ref{fig: qs}).  Thus, as $s$ decreases towards the free end of
the A block at $s_0$, $q^\dag(\textbf{r},s)$ assumes the
probability of all the higher generations diffusing ``into" this
lower generation branch.  We will refer to this diffusion from
$s_g$ towards $s_0$ as ``backward motion."  Note that in terms of
the probability distributions all paths diffusing from any of the
$\eta_g$ free ends are equivalent, and therefore,
$q^\dag(\textbf{r},s)$ is well defined.

We summarize the above definition by writing
$q^\dag(\textbf{r},s)$ in terms of our unbranched propagator,
$G (\textbf{r}_i, s_i ; \textbf{r}_f, s_f) $:
\begin{multline}
q^\dag(\textbf{r},s) = \int d \textbf{r}_{\alpha} \
G (\textbf{r}, s ; \textbf{r}_{\alpha},
s_{\alpha}) \ [ q^\dag(\textbf{r}_\alpha,s_\alpha^+)
]^{\eta_{\alpha+1}} \ , \\ \textrm{for  } s_{\alpha-1} > s >
s_{\alpha} \ ,
\end{multline}
where $q^\dag(\textbf{r}_\alpha ,s_\alpha^+)$ indicates that we
take the value of this function from the end of the higher
generation at $s_\alpha$ (just before the branch point).  If we
normalize our propagator so that $\lim_{s_f \rightarrow s_i}
G (\textbf{r}_i, s_i ; \textbf{r}_f, s_f) =
\delta(\textbf{r}_f - \textbf{r}_i)$, then we establish a set of
boundary conditions for $q^\dag(\textbf{r},s)$ at its free end at
$s_g$ and each branching point,
\begin{equation}
\label{eq: qd(sg)} q^\dag(\textbf{r},s_g^-) = 1 \ ,
\end{equation}
\begin{equation}
\label{eq: qd(sa)} q^\dag(\textbf{r},s_\alpha^-) = [
q^\dag(\textbf{r}_\alpha,s_{\alpha}^+) ]^{\eta_{\alpha+1}} \ ,
\end{equation}
where $q^\dag(\textbf{r},s_\alpha^-)$ is the limit of the function
as $s$ approaches $s_\alpha$ from below (just after the branching
point).  Thus, at a given branching point, $s_\alpha$ the value of
$q^\dag(\textbf{r},s)$ changes discontinuously, from
$q^\dag(\textbf{r},s_\alpha^+)$ to
$q^\dag(\textbf{r},s_\alpha^-)$, since the function assumes
the probability of the other higher generation branches meeting it
at that point.

Because $q^\dag(\textbf{r},s)$ is defined in terms of the
propagator, $G (\textbf{r}_i, s_i ;
\textbf{r}_f, s_f)$, we know that it will obey the same diffusion
equation as the propagator.  Namely,
\begin{equation}
\label{eq: dqds} - \frac{\partial q^\dag}{\partial s} =
\left\{\begin{array}{ll} \frac{Na^2}{6} {\mathbf{\nabla}}^2 q^\dag
- w_A(\textbf{r})q^\dag \ ,
& \textrm{for $s_0 < s < s_1$} \ , \\ \\
\frac{Na^2}{6 \kappa^2} {\nabla}^2 q^\dag - w_B(\textbf{r})q^\dag \ ,  &
\textrm{for $s_1 < s < s_g$} \ ,
\end{array}\right.
\end{equation}
It should be understood that we will solve these first-order
equations for the unbranched portions of the chain and use the
branching points to determine boundary conditions; hence, we do not need
to worry about differentiating at branching points.

Because Eq. (\ref{eq: dqds}) is a linear equation for $q^\dag$ which is first order in $s$, given any set
of fields, $w_A(\mathbf{r})$ and $w_B(\mathbf{r})$, we can solve
for $q^\dag(\textbf{r},s)$ for all segments.  First, using
(\ref{eq: dqds}) and (\ref{eq: qd(sg)})  we solve for the
$q^\dag(\textbf{r},s)$ for the $g^{\rm{th}}$ generation.  Then, we
can use (\ref{eq: dqds}), (\ref{eq: qd(sa)}) and our solution for
$q^\dag(\textbf{r},s_{g-1})$ to solve for the $(g-1)^{\rm{th}}$
generation of $q^\dag(\textbf{r},s)$.  Likewise, we can then
iteratively solve for all lower generations until we get to the
first.

Once the value of $q^\dag(\textbf{r},s)$ is known for all $s$ down
to $s_0$, we can compute the single-chain partition by integrating
this backward motion end-distribution function over the position
of the free end of the A-block,
\begin{equation}
\mathcal{Q} = \int d \textbf{r} \ q^\dag(\textbf{r},s_0) \ .
\end{equation}
However, in order to compute the mean-field melt free energy we
need to calculate the average monomer distributions,
$\phi_A(\textbf{r})$ and $\phi_B(\textbf{r})$, created by the
monomer potentials, $w_A(\mathbf{r})$ and $w_B(\mathbf{r})$. By
introducing another end-distribution function, $q(\textbf{r},s)$,
we can compute the functional derivative of $- \ln \mathcal{Q}$
with respect to these fields directly.

We define $q(\textbf{r},s)$ to be proportional to probability that
a chain configurations diffuses in the ``forward" direction from
its other free end (the free end of the $A$ block at $s_0$) along
one of the branched trajectories of the molecule to $s$ at the
position $\textbf{r}$ (see Fig. ~\ref{fig: qs}).  At branching
points, $s_\alpha$, $q(\textbf{r},s)$ assumes the probability that
$(\eta_{\alpha+1} - 1 )$ branches have also diffused from their
free ends at $s_g$ to $\textbf{r}_\alpha$ at $s_\alpha$. This is
to say that $q(\textbf{r}_\alpha,s_\alpha^+)$ contains not only
the probability that the $s_0$ end diffused to this point but also
the probability that all of the other branches, not including the
currently diffusing path, have diffused to to $\textbf{r}_\alpha$
at $s_\alpha^+$ to meet it.  This property makes $q(\textbf{r},s)$
convenient for computing the average monomer distributions. Using
the above definition we have,
\begin{multline}
q(\textbf{r},s) = \int \! d \textbf{r}_{\alpha}
G (\textbf{r}_{\alpha}, s_{\alpha}; \textbf{r},
s) \ q(\textbf{r}_\alpha,s_\alpha^-) \ [
q^\dag(\textbf{r}_\alpha,s_\alpha^+) ]^{\eta_{\alpha}-1}  , \qquad
\\ \textrm{for  } s_{\alpha} > s > s_{\alpha+1} \ .
\end{multline}
The corresponding boundary conditions for $q(\textbf{r},s)$ are
given by,
\begin{equation}
\label{eq: q(s0)}
q(\textbf{r},s_0^+) = 1 \ ,
\end{equation}
\begin{equation}
\label{eq: q(sa)}
q(\textbf{r},s_\alpha^+) = q(\textbf{r},s_\alpha^-) \ [
q^\dag(\textbf{r}_\alpha,s_{\alpha}^+) ]^{\eta_{\alpha}-1} \ .
\end{equation}
Since the ``motion'' of the diffusion along the chain is reversed
from that of $q^\dag(\textbf{r},s)$ the diffusion equation for
$q(\textbf{r},s)$ is the same as (\ref{eq: dqds}) except with
a plus sign appearing on the left hand side.
In analogy with $q^\dag(\textbf{r},s)$, we must first solve the
diffusion and (\ref{eq: q(s0)}) for the first generation of
$q(\textbf{r},s)$.  We then use our second generation solution of
$q^\dag(\textbf{r},s)$ and the first generation solution of
$q(\textbf{r},s)$ in (\ref{eq: q(sa)}) to find the solution for
the second generation.  We can repeat the process to solve for
$q(\textbf{r},s)$ over the entire length from $s_0$ to $s_g$.

It is not difficult to show that the monomer distributions, given by
Eq. (\ref{eq: pAB}), can be computed by,
\begin{equation}
\label{eq: phiAreal} \phi_{A}(\textbf{r}) = \frac{V}{\mathcal{Q}}
\int_{s_0}^{s_1} \!\! ds \ q(\textbf{r},s) q^\dag(\textbf{r},s) \
,
\end{equation}
\begin{equation}
\label{eq: phiBreal} \phi_{B}(\textbf{r}) = \frac{V}{\mathcal{Q}}
\sum_{\alpha=2}^{g} \mathcal{N}_{\rm{B},\alpha}
\int_{s_{\alpha-1}}^{s_\alpha} \!\!\! ds \ q(\textbf{r},s)
q^\dag(\textbf{r},s) \ ,
\end{equation}
where $V = \frac{n N}{\rho_0}$ and
$\mathcal{N}_{\rm{B},\alpha}$ is the number of B-blocks in the
$\alpha^{\rm{th}}$ generation, which is simply given by
$\eta_\alpha \eta_{\alpha-1} \ldots \eta_2$. Thus, the mean-field
free energy, (\ref{eq: Freal}), can be computed entirely with the
end-distribution functions, $q(\textbf{r},s)$ and
$q^\dag(\textbf{r},s)$.

While real-space methods for numerically solving these diffusion
equations exist, \cite{fred_prl_99, fred_macro_review_02} these methods tend to be
computationally intensive for melt phases with spatial variation
in three dimensions.  Instead, we use Fourier expansions of the functions to solve for $q^\dag(\textbf{r},s)$ and
$q(\textbf{r},s)$ given an arbitrary set of external fields,
$w_A(\mathbf{r})$ and $w_B(\mathbf{r})$.  Since we know that
equilibrium structures are themselves infinitely periodic
structures, we expect that we can very accurately describe
mean-field results with a finite number of Fourier terms included
the expansion.  For up to moderately large degrees of segregation (for $\chi N \lesssim 50$) the spectral methods of \cite{matsen_schick_prl_94} and \cite{matsen_schick_macro_94} allow for the rapid and very accurate exploration of mean-field thermodynamics \cite{fred_macro_review_02}.  We present the spectral form of our SCFT for multiply-branched copolymer melts in the Appendix.

\section{Doubly Functional Branching: The Role of Interfaces}

\begin{figure}\center
\epsfig{file=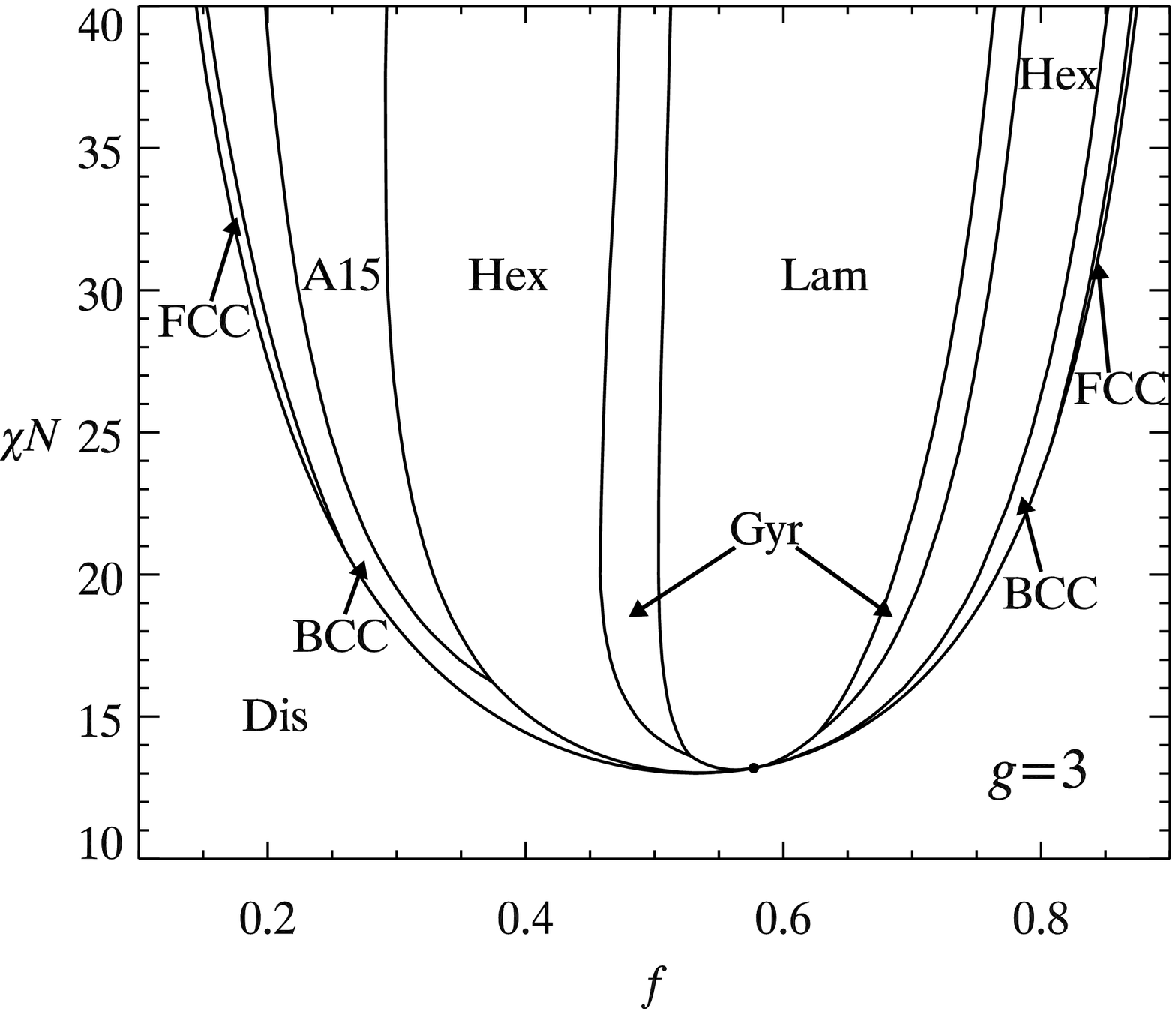,width=3.0in}\center
\epsfig{file=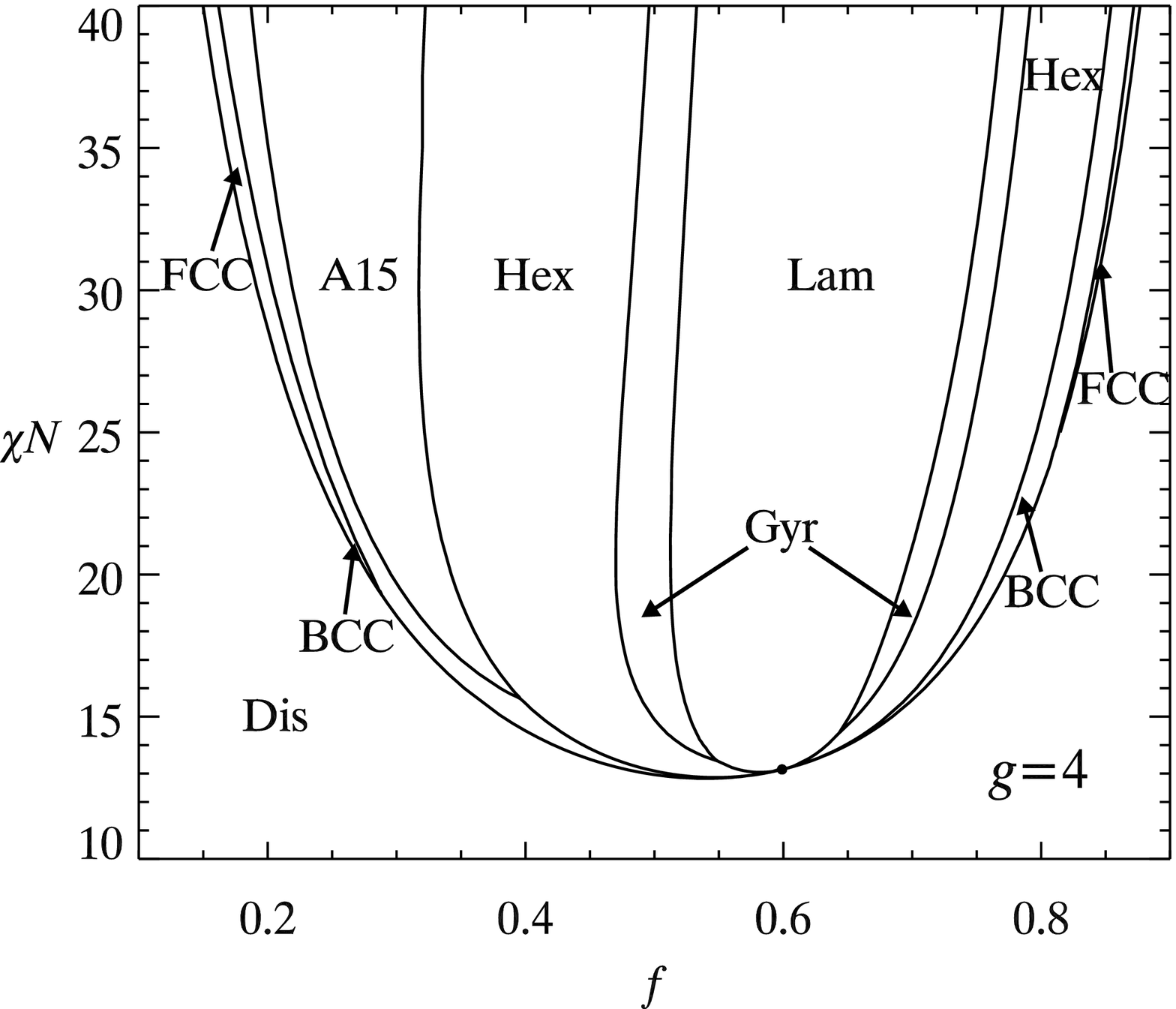,width=3.0in}\center
\caption{Phase diagrams for $g=3$ and $g=4$. Dis labels regions where the melt is disordered. Stable regions of ordered phases are labeled:  (Lam) lamellar;  (Gyr) gyroid, $Ia\bar{3}d$ symmetry; (Hex) hexagonal-columnar, $p6mm$ symmetry; (A15) sphere phase, $Pm\bar{3}n$ symmetry; (BCC) body-center cubic lattice of spheres, $Im\bar{3}m$ symmetry; and (FCC) face-centered cubic lattice of spheres, $Fm\bar{3}m$ symmetry \cite{FCC_note}.  The circle marks the mean field critical point through which the system can transition from the disordered state to the Lam phase via a continuous, second-order phase transition.  All other phase transitions are first-order} \label{fig: phase34}
\end{figure}

\begin{figure}\center
\epsfig{file=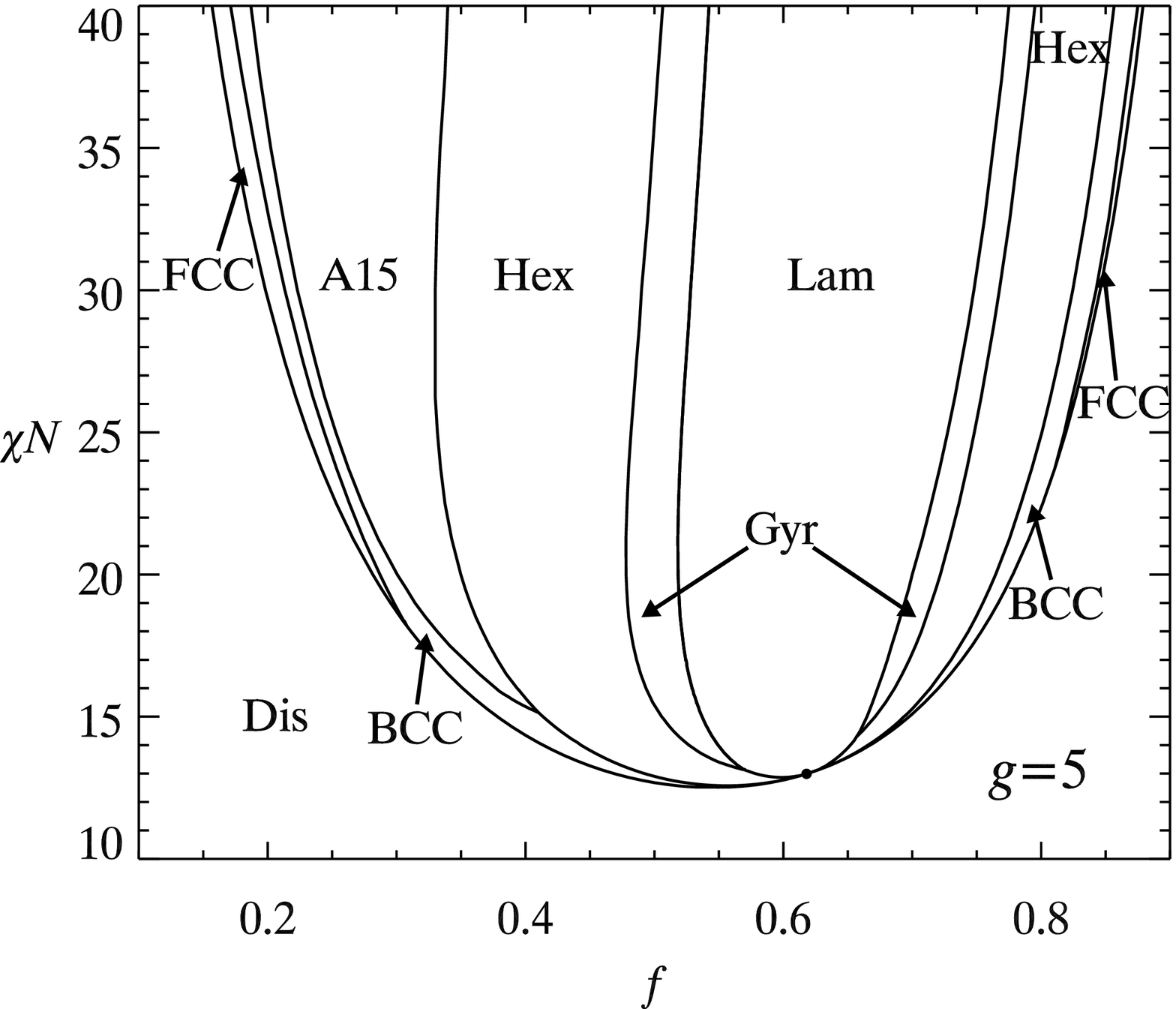,width=3.0in}\center
\epsfig{file=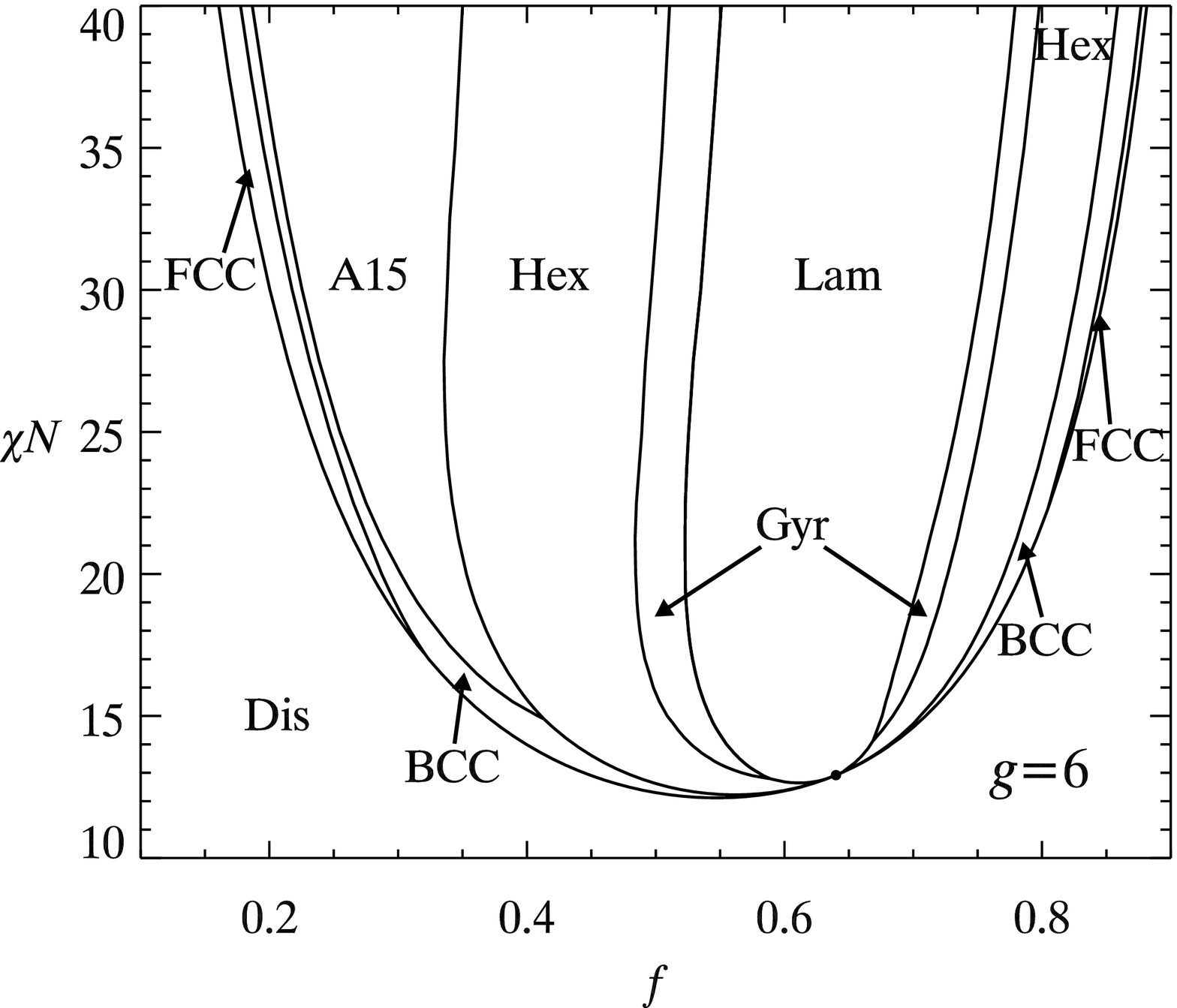,width=3.0in}\center
\caption{Phase diagrams for $g=5$ and $g=6$. Labels appear as in Figure \ref{fig: phase34}.} \label{fig: phase56}
\end{figure}

Using the SCFT derived in the previous section we computed the $\chi N \leq 40$ mean-field phase behavior for multiply-branched copolymer melts where the branching, or functionality, of each generation is 2.  We compute the phase behavior for $g=2\ldots6$ for monomers of equal segment size, $\kappa =1$.  To achieve a precision of $\pm 10^{-3}$ in $f$ and $\pm 10^{-2}$ in $\chi N$ we require a precision in the free energy of better than $\pm 10^{-4}$.  This requires the use of up to 908 basis functions for some phases.  The mean-field phase diagrams for $3 \leq g \leq 6$ are shown in Figs. \ref{fig: phase34} and \ref{fig: phase56}.  We have already reported on the phase behavior for $g=2$, the AB$_2$ miktoarm star \cite{grason_kamien_macro_04}.  

\begin{figure}\center
\epsfig{file=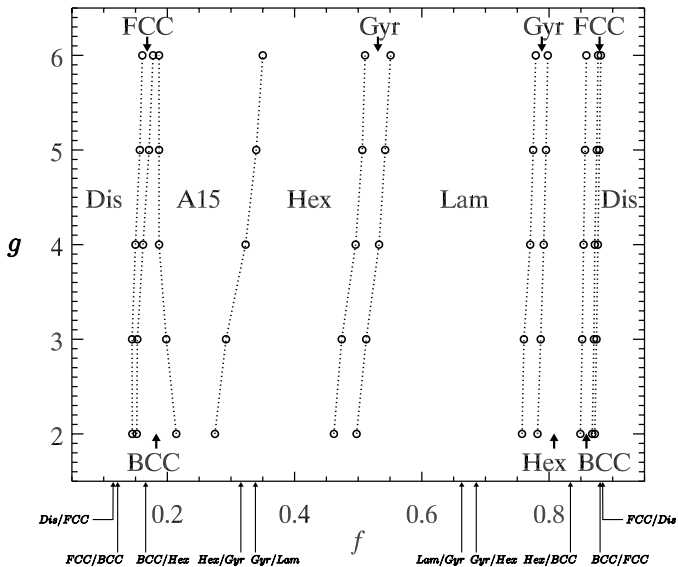,width=3.5in}\center
\caption{The SCFT phase boundaries computed at $\chi N =40$ for $2\leq g \leq 6$ are depicted as open circles.  For comparison the $\chi N = 40$ phase boundaries for linear diblocks are indicated of the $f$ axis.  Note the absence of a stable A15 phase for linear AB diblock copolymer melts.} \label{fig: branchedphase}
\end{figure}

The thermodynamics of these melts are strongly influenced by the introduction of the multiply-branched architecture.  Compared to the predicted phase behavior of linear AB block copolymer melts, the phase boundaries of these branched copolymer melts are skewed systematically towards larger values of $f$ for most phases \cite{matsen_schick_prl_94}.  This indicates an enhanced preference for phases where the branched polymer domain is on the convex side of curved AB interfaces.  In Figure \ref{fig: branchedphase} we plot the strong-segregation ($\chi N =40$) phase boundaries as a function of branching generation.  The preference for morphologies with the branched, B domain on the outside of a highly curved interface is increases with increasing generation.  For example, spherical micelles where the A blocks form the core region are stable up to $f=0.275$ for $g=2$ but stable up to $f=0.350$ for $g=6$.  This effect is well established for copolymer architectures with elastically asymmetric blocks \cite{olmsted_milner_macro_98, vavasour_whitmore_macro_93, matsen_schick_macro_94, matsen_bates_jpolysci_97}.

In general, elastic asymmetry stems chiefly from two sources---asymmetric monomer sizes and asymmetric copolymer architecture.  Milner demonstrated within SST that the elastic asymmetry between copolymer blocks of a A$_n$B$_m$ miktoarm star is captured by the parameter, $\epsilon=\frac{n}{m} (\frac{\rho_B a_B^2}{\rho_A a_A^2})^{1/2}$, where $\rho_A^{-1}$ and $\rho_B^{-1}$ are the respective volumes of the A and B segments \cite{milner_macro_94}.  From this analysis it can be shown that the effective spring constant of the B brush domain is a factor of $\epsilon^2$ times the value of the symmetric case (for $\epsilon=1$).  For $\epsilon > 1$, the molecular asymmetry leads to the stabilization of morphologies where the B polymer block composes the outer corona of spherical and cylindrical domains for larger values of A composition than is observed for elastically symmetric copolymers \cite{yang_gido_macro_01}.

It is desirable to have a similar quantitative measure of the elastic asymmetry for copolymers with this multiply-branched structural motif.  However, in contrast to the miktoarm star architecture, the elastic enhancement of multiply-branched domains depends on the aggregate morphology.  Using the Alexander-de Gennes, strong-segregation analysis employed by Frischknecht and Fredrickson we find, for example, that the stiffness of a lamellar B domain in these doubly-functional copolymer melts is enhanced by a factor of $ 4(8^{g-1}-1)/[7(2^{g-1}-1)]$ over the linear, unbranched case \cite{fred_macro_99}.  This corresponds to factors of 4, 12, $\frac{292}{7}\simeq 41.7$, 156 and 604 multiplying the stretching free energy of a lamellar B domain for the $g=2, 3, 4, 5$ and 6 case respectively.  Pickett demonstrates, however, that the Alexander-de Gennes approximation provides an overestimate of the branched chain free energy whose error grows quickly with the branching generation \cite{pickett_macro_02}.  Based on the analysis of Pickett \cite{pickett_macro_02} for a slightly different copolymer architecture we might expect that by relaxing the constraint that the chain ends are held at the tips of the brush, the SST stretching free energy of the branched B domain can by relaxed from the Alexander-de Gennes upper limit by factors of roughly 2.6, 5.5, 11.7 and 24.6 for $g=3, 4, 5$ and 6, respectively (the $g=2$ case corresponds the AB$_2$ miktoarm star).  This allows us to estimate more realistic values of the elastic asymmetry in the lamellar morphology:  4 for $g=2$; 4.6 for $g=3$; 7.7 for $g=4$; 13.4 for $g=5$; and 24.6 for $g=6$.  While these are somewhat crude estimates, they provide reasonable correspondence between the SST phase behavior of these multiply-branched copolymers and AB$_n$ copolymers \cite{grason_kamien_macro_04}.  For example, the respective $\chi N =40$,  Gyr-Lam and A15-Hex transitions occur at $f=0.546$ and $f=0.349$ for a melt of AB$_5$ miktoarm stars, corresponding to an elastic asymmetry of 25.  This should be compared to the same transitions which occur at $f=0.550$ and $f=0.350$, respectively, for a $g=6$ branched copolymer, with an estimated elastic asymmetry of 24.6.

We note the stability of the cubic A15 phase in these melts.   We have argued \cite{grason_didonna_kamien_prl_03} that as $\chi N \rightarrow \infty$ and in the limit that the AB interface of a sphere phase is constrained to adopt the same shape as the lattice unit cell that the A15 should be
the equilibrium structure.  In this limit the relative stability of competing arrangements of micelles can be assessed purely in terms of two geometric moments of the Voronoi polyhedra of the lattice: the area and the second-moment of the lattice .  If $R_X(\hat{\Omega})$ measures the radial distance from the center to the surface of the Voronoi cell of lattice $X$ at solid angle, $\hat{\Omega}$, then we can compute the area in terms of the area of a spherical cell of equal volume,
\begin{equation}
\mathcal{A}_X = \frac{1}{(4 \pi)^{1/3}}\frac{\int d \hat{\Omega} \big[R^2_X (\hat{\Omega})+(\nabla_{ \hat{\Omega}} R_X (\hat{\Omega}))^2 \big]}{ \big(\int d \hat{\Omega} R^3_X (\hat{\Omega}) \big)^{2/3}} \ ,
\end{equation}
where $\nabla_{\hat{\Omega}}=\hat{\theta} \frac{\partial}{\partial \theta} + \hat{\phi} \frac{\partial}{\partial \phi}$.  We can also define the second-moment, or ``stretching" moment, of the Voronoi cell,
\begin{equation}
\mathcal{I}_X = (4 \pi)^{2/3} \frac{\int d \hat{\Omega} R^5_X (\hat{\Omega})}{ \big(\int d \hat{\Omega} R^3_X (\hat{\Omega}) \big)^{5/3}} \ ,
\end{equation}
where again we have normalized by the same measure for a spherical cell of equal volume.  It can be shown that the free energy per chain in a micellar phase arranged in lattice $X$ is simply given by $F_X = F_0 (\mathcal{A}^2 \mathcal{I})^{1/3}$, where $F_0$ is the free energy per chain for the case when the Voronoi cell is approximated as a sphere \cite{olmsted_milner_macro_98, grason_didonna_kamien_prl_03}.  Given these geometric measures for all candidate arrangements of spherical micelles we can assess the relative stability of these phases in this limit where AB interface has the same shape as the unit cell of the lattice.  It was discovered by Weaire and Phelan that the space partition of the A15 lattice has the lowest area of all known equal volume periodic partitions of three dimensional space \cite{weaire_phelan_94}.  It is for this reason, despite the fact that the BCC lattice has a smaller second moment, that the A15 lattice is most stable among the lattice arrangements of spherical micelles when AB interfaces have adopted the shape of the Voronoi cell in which they are confined.  In particular, this limit predicts that the free energy per chain for the A15 phase is 0.14\% and 0.61\% lower than the BCC and FCC phases, respectively.  Of course, there are finite $\chi N$ corrections to this asymptotic limit due to chain fluctuations which are neglected in the strong-segregation limit, but the lowest order corrections which distinguish between morphologies are smaller than the leading order free energy term by a factor of $(\chi N)^{-4/9}$ \cite{goveas_milner_russel_macro_97, likhtman_semenov_epl_00}.  

The conclusions of our SST analysis are valid in the limit that the AB interface has adopted the polyhedral shape of the lattice Voronoi cell.  It is well known that constraining a micelle to occupy a polyhedral unit cell frustrates the internal configuration of the aggregate \cite{matsen_bates_macro_96, fredrickson_macro_93}.  Chains which extend along directions towards corners of the Voronoi cell must stretch further than those extending towards the walls.  The difference in tension in these chains leads to a tendency to distort the AB interface from its ideal, uniformly curved shape into the polyhedral shape of the Voronoi cell.  Of course, the micelle will adopt some compromise between the uniform curvature and relaxed outer chain stretching which will be determined by the relative importance of outer chain stretching and the forces which pull inward on the AB interface, namely the surface tension and the inner chain stretching.  We demonstrated \cite{grason_kamien_macro_04} how the tendency for cylindrical micelles in the Hex phase to adopt an hexagonal interface shape is enhanced both by an increase in inner domain volume fraction, $f$, and the elastic asymmetry between the coronal and core polymer domains, $\epsilon$.  In particular, we found that although AB interfaces in micelles for symmetric molecules (e.g. linear AB diblocks) remain relatively unperturbed by the lattice symmetry, cylindrical micelles composed of very elastically asymmetric copolymers ($\epsilon \gtrsim 3$) have interfaces which are very nearly hexagonal in regions where the Hex phase is thermodynamically stable.  Given the stability of the A15 phase in the present system (Figure \ref{fig: branchedphase}), it must be that the interfaces of the sphere phases are also substantially distorted by the polyhedral environment of the lattice Voronoi cell.

\begin{figure}\center
\epsfig{file=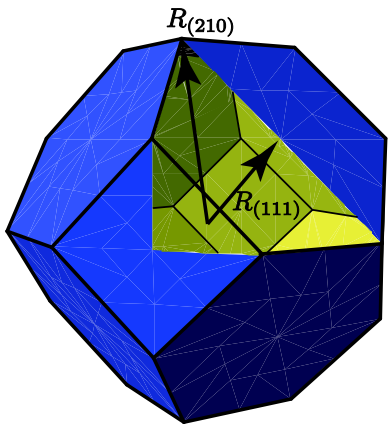 ,width=2.0in}\center
\caption{A view of the Voronoi cell of the BCC lattice, a truncated octahedron, with half of one hexagonal face and a quarter of one square face removed to reveal the inside.  The edges are drawn as black lines, the outside is shown as blue and the inside is shown as yellow.  The vectors connecting the center of the cell to the corners along the (210) directions and the nearest walls along the (111) directions are shown. } \label{fig: cutaway}
\end{figure}

\begin{figure}\center
\epsfig{file=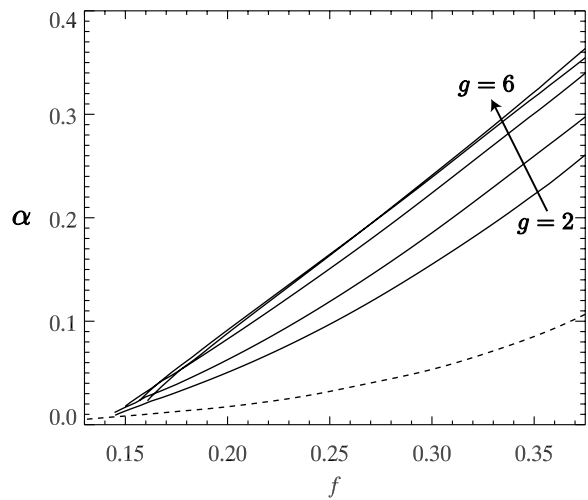 ,width=3.0in}\center
\caption{Plot of the measured distortion, $\alpha$ as defined in eqs. (\ref{eq: delta}) and (\ref{eq: alpha}), measured from SCFT results for the BCC phase of branched copolymer melts for generations, $2\leq g \leq6$. For comparison the dashed line shows the same distortion for linear AB diblock copolymer melts.}
\label{fig: alpha}
\end{figure}

We can quantify the extent to which AB interfaces in sphere phases adopt the shape of the Voronoi cells from our SCFT results.  A measure of the distortion of the interface of a micelle in the BCC phase from the ideal spherical shape is the difference of the distances from the center of the micelle to the interface along directions towards the closest face of the Voronoi cell,  the (111) direction, and towards the corner of the Voronoi cell,  the (210) direction,
 \begin{equation}
 \label{eq: delta}
\delta \equiv \frac{R_{(210)}-R_{(111)}}{R_{(210)}+R_{(111)}} \ ,
\end{equation}
where $R_{(210)}$ and $R_{(111)}$ are the radial distances to the AB interface along those directions (see Figure \ref{fig: cutaway}).  For a spherical interface we have $\delta_{sph}=0$ and for an interface which as the truncated-octahedron shape of the BCC Voronoi cell, $\delta_{BCC}=(\sqrt{5}-\sqrt{3})/(\sqrt{5}+\sqrt{3}) \simeq 0.127$.  By normalizing measured values of $\delta$ by $\delta_{BCC}$, we can assess the polyhedral distortion on the scale set by the shape of the BCC Voronoi cell.  Therefore, we use,
\begin{equation}
\label{eq: alpha}
\alpha \equiv \frac{\delta}{\delta_{BCC}}
\end{equation}
 to quantify the polyhedral distortion of the interface as a function of molecular architecture.  Figure \ref{fig: alpha} plots the shape parameter, $\alpha$, measured from SFCT calculations for the BCC phase as function of $f$ and branching generation $g$.  It is clear that the packing frustration introduced by the polyhedral Voronoi cell increases as the volume fraction of the core of the micelle grows.  Although the close-packing limit of hard spheres in a BCC lattice is at a volume fraction of $\sqrt{3} \pi/8 \simeq 0.68024$, the cores of the micelles are highly deformed for $f$ well below this.  Since the outer chain stretching is responsible for this polyhedral distortion, the tendency to adopt the truncated-octahedral shape of the BCC lattice is enhanced as the stiffness of the coronal region is increased by molecular branching.

 \begin{figure*}\center
\epsfig{file=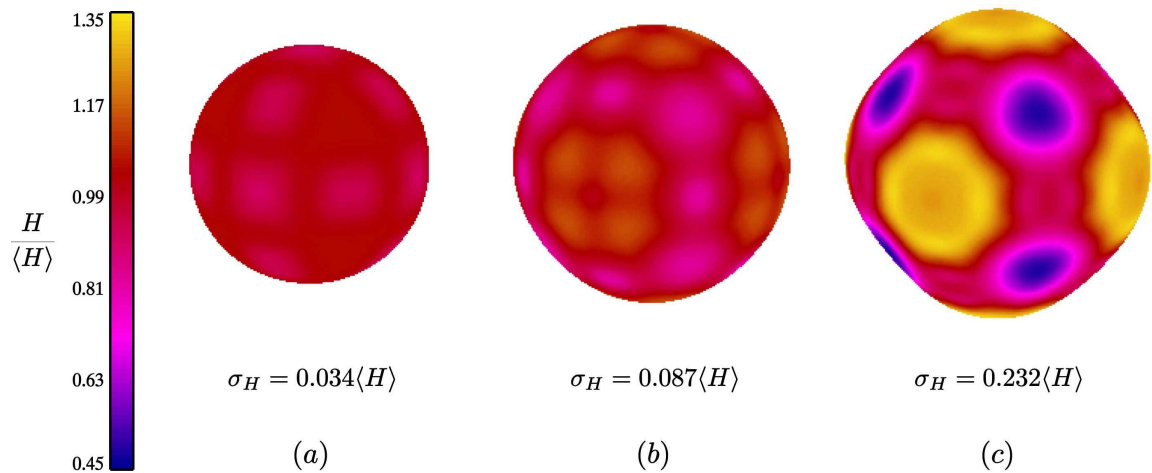 ,width=6.0in}\center
\caption{The AB interfaces extracted from the SCFT calculation (the surfaces at which $\phi_A ({\bf r})=0.5$) for the BCC phase along the thermodynamic phase boundary separating spherical and cylindrical morphologies:  (a) for linear diblocks at $f=0.166$; (b) for $g=2$ branched copolymers at $f=0.275$; and (c) for $g=6$ branched copolymers at $f=0.350$.  The interfacial distortion at these points corresponds to measured values of $\alpha= 0.011$, 0.124 and 0.321, respectively.  The surfaces are shaded according to the local mean curvature, $H$, measured in units of the average mean-curvature, $\langle H \rangle$.  The variation of the mean-curvature provides a direct measure of the variation of the polymer chain tension at the interface, due to the polyhedral environment of lattice Voronoi cell.  The standard deviation of the curvature, $\sigma_H$, for each surface is given in units of $\langle H \rangle$.  } \label{fig: curvature}
\end{figure*}

To further visualize this distortion we compute the mean-curvature, $H$, of AB interfaces extracted from SCFT results for melts at $\chi N =40$ at the phases boundary between spherical and cylindrical phasess, the A15-Hex boundary, for $g=6$, $g=2$, and for linear diblocks.  The distortion in these interfaces corresponds to measured values of $\alpha=0.32$, $\alpha=0.128$ and $\alpha=0.011$, respectively.   These surfaces are displayed in Figure \ref{fig: curvature}, and following the analysis of Matsen and Bates \cite{matsen_bates_macro_96} the interfaces are shaded according to the local mean-curvature.  As the polyhedral distortion increases, the deviation from constant mean curvature grows.  The surface tension, $\gamma$, associated with an interface between unlike polymer melt domain scales as $\chi^{1/2}$ \cite{semenov_jetp_85}.  A patch of area, $dA$, of a curved interface experiences a force due the surface tension which is given by $2 H \gamma dA$ \cite{safran_book}.  Since these micelle configurations are saddle-points of the free energy, we know that the force due to the tension pulling inward on the interface is balanced by a net force pulling outward on the micelle interface, that is, the interface must be in mechanical equilibrium.  In copolymer micelles the compensating forces are due to a difference in the tension of the chains in the core and coronal domains.  Therefore, variation of the mean curvature of the interface provides a direct measure of the chain tension pulling on the interface.  Regions of high interfacial curvature, towards the edges and vertices of the BCC Voronoi polyhedron, correspond to regions where the coronal B domains pull relatively strongly on the interface.  Conversely, relatively flat regions on the AB interface, towards the nearest-neighbor faces of the polyhedron, indicate that the chain stretching is relatively low.

\section{Conclusion}
From our analysis we see that for elastically asymmetric copolymers the polyhedral shape of the lattice Voronoi cell forces the micelle configuration to deviate drastically from the limit of uniform interfacial curvature.  While we have argued that in the limit of perfectly polyhedral interfaces the A15 phase is most stable among the sphere phases and that very elastically asymmetric micelles approach this polyhedral limit in regions where sphere phases are thermodynamically stable, it has yet to be shown that for small distortions (for $\alpha \lesssim 0.35$) the minimal Voronoi cell area argument should apply.  For example, the reduced area of the $g=6$, BCC interface of Figure \ref{fig: curvature} is $1.0094$, to be compared to the reduced area of the truncated-octahedron of the BCC Voronoi cell, $1.0990$.  In this sense, AB interfaces of physical micelles seems to be distorted less than about 10\% towards their Voronoi polyhedra.  Nevertheless, we argue that the polyhedral interface limit of the micelle configurations sets the scale of the frustration.  While the true micelle interfaces are some interpolation between a spherical and polyhedral shape, the scale of the frustrated free energy is set by the polyhedral interface upper bound.  As mentioned above, in the limit of polyhedral interfaces as $\chi N \rightarrow \infty$, SST predicts that $F_{BCC}=1.0014 F_{A15}$ and $F_{FCC}=1.0061 F_{A15}$.  We can compare this prediction to the results of our SCFT calculations along the  Hex-A15 phase boundary at $\chi N =40$ for $g=5$ branched copolymers (at $f=0.340$) for which we find $F_{A15}=6.296 n k_BT$, $F_{BCC}=6.314 n k_B T$ and $F_{FCC}=6.326 n k_B T$, corresponding to 0.28\% and 0.46\% higher free energy than the A15 phase for BCC and FCC phases, respectively.  On the scale of these small free energy differences, the analysis of our geometrical limit is a necessary component of any rational explanation for lattice choice.  Therefore, while such a calculation has yet be carried out, we expect that a more detailed SST analysis of the relaxed configurations of micelle interfaces will bear this argument out.

\section*{Acknowledgments}
It is a pleasure to acknowledge stimulating discussions with D. Discher, B. DiDonna, P. Heiney, and V. Percec.  This work
was supported by NSF Grants DMR01-02459, DMR01-29804, and INT99-10017,  a gift from Lawrence J. Bernstein, and the
Pennsylvania Nanotechnology Institute.
\appendix*

\section{Spectral SCFT}

Following Matsen and Schick \cite{matsen_schick_prl_94,matsen_schick_macro_94} we
define a set of orthogonal basis functions, $f_i(\textbf{r})$, which have the
periodic symmetry of our copolymer phase.  We expand all of the necessary functions of position in this basis, so that $g(r)=\sum_i g_i f_i({\bf r})$.  For example, the BCC
phase of spheres can be described by the set of functions with
$Im\bar{3}m$ symmetry.  The functions are normalized such that,
\begin{equation}
V^{-1} \int d \textbf{r}\  f_i(\textbf{r}) f_j(\textbf{r}) =
\delta_{ij} \ .
\end{equation}
In addition, we demand that these functions are eigenfunctions of
the Laplacian operator so that,
\begin{equation}
\nabla^2 f_i(\textbf{r}) = - \frac{\lambda_i }{D^2}
f_i(\textbf{r}) \ ,
\end{equation}
where $D$ is the length scale of the periodicity of the system.
The set of functions is ordered in an increasing
sequence in $\lambda_i$, and $\lambda_1$ is set to 0 (or
$f_1(\textbf{r})=1$).  Because the product of two basis functions,
$f_i(\textbf{r}) f_j(\textbf{r})$, has the all symmetries of the
basis, it belongs in the same space of functions as our basis.
Thus, we can write the product as expansion in our basis
functions.  We define the coefficients, $\Gamma_{ijk}$, of the
expansion so that,
\begin{equation}
\label{eq: Gamma} f_i(\textbf{r}) f_j(\textbf{r}) = \sum_k
\Gamma_{ijk}  f_k(\textbf{r})
  \ .
\end{equation}
Alternately, given the set of basis functions invariant under all of the symmetry operations of the group, this coefficient
can be computed by $\Gamma_{ijk} = V^{-1} \int d\textbf{r}
f_i(\textbf{r}) f_j(\textbf{r}) f_k(\textbf{r})$.

With these definitions and a finite Fourier expansion of all
functions of position we can rewrite the diffusion equation, (17),
as a matrix equation,
\begin{equation}
\label{eq: dqids}
 - \frac{\partial q_i^\dag}{\partial s} = \left\{
\begin{array}{ll}
 \sum_{j} A_{ij} q_j^\dag , & \textrm{for $s_0 < s < s_1$} \ , \\ \\
 \sum_{j} B_{ij} q_j^\dag , & \textrm{for $s_1 < s < s_g$} \ ,
\end{array}\right.
\end{equation}
where we have defined the matrices, $A_{ij}$ and $B_{ij}$,
\begin{equation}
A_{ij} \equiv  -\frac{Na^2 \lambda_i}{6 D^2} \delta_{ij} - \sum_k
w_{A,k}
 \Gamma_{ijk} \ ,
\end{equation}
\begin{equation}
B_{ij}  \equiv  -\frac{Na^2 \lambda_i}{6 \kappa^2 D^2} \delta_{ij}
- \sum_k w_{B,k}  \Gamma_{ijk} \ .
\end{equation}
Since these are symmetric, real matrices we can diagonalize
$A_{ij}$ and $B_{ij}$ by orthogonal transformations, such that,
$\sum_{k,l} [O_A^T]_{ik} A_{kl}  [O_A]_{lj} = \mathcal{A}_i
\delta_{ij}$ and $\sum_{k,l} [O_B^T]_{ik} B_{kl} [O_B]_{lj} =
\mathcal{B}_i \delta_{ij}$, where $\mathcal{A}_i$ and
$\mathcal{B}_i$ are the eigenvalues of $A_{ij}$ and $B_{ij}$, and
$[O_A]_{ij}$ and $[O_B]_{ij}$ are the orthogonal matrices which
diagonalize, $A_{ij}$ and $B_{ij}$, respectively.  The matrix,
\begin{equation}
T_{\rm{A},ij}^{\dag} (s'-s) \equiv \sum_k [O_A]_{ik} \exp \{-
\mathcal{A}_k (s' - s) \} [O_A]_{jk} \ ,
\end{equation}
transfers the A-block solution to (\ref{eq: dqids}) from $s$ to
$s'$, and the matrix $T_{\rm{B},ij}^{\dag} (s'-s)$ does the same
for the B-block solution.  Using these matrices we can write the
solutions for $q_i^{\dag} (s)$,
\begin{equation}
\label{eq: qdi(s)} q_i^\dag (s) = \left\{\begin{array}{ll}
\sum_{j} T_{\rm{A},ij}^{\dag} (s-s_1)
\Lambda_j^\dag (s_1) \ , & \textrm{for $s_0 < s < s_1$} \ , \\ \\
\sum_{j} T_{\rm{B},ij}^{\dag} (s-s_\alpha) \Lambda_j^\dag
(s_\alpha) \ , & \textrm{for $s_{\alpha - 1} < s < s_\alpha$} \ ,
\\ (\alpha \neq 1) \ ,
\end{array}\right.
\end{equation}
where $\Lambda_i^\dag (s_\alpha)$ are the boundary conditions for
$q_i^\dag (s)$ at $s_\alpha^-$:
\begin{equation}
\Lambda_i^\dag (s_\alpha) = V^{-1} \int d\textbf{r} \ [q^\dag
(\textbf{r}, s_\alpha^+)]^{\eta_{\alpha+1}} f_i(\textbf{r}).
\end{equation}
Thus, we have that $\Lambda_i^\dag (s_g) = \delta_{i1}$.

In order to compute $\Lambda_i^\dag (s_\alpha)$ for the lower
generations, we define the function,
\begin{equation}
\psi_i^{(m)} (s_\alpha) \equiv V^{-1} \int d\textbf{r} \ [q^\dag
(\textbf{r}, s_\alpha^+)]^m f_i(\textbf{r}).
\end{equation}
Given this definition we have $\psi_i^{(1)} (s_\alpha) = q_i^\dag
(s_\alpha^+)$ and, of course, $\Lambda_i^\dag (s_\alpha) =
\psi_i^{(\eta_{\alpha+1})} (s_\alpha)$. Using (\ref{eq: Gamma})
and the fact that the Fourier expansion of $q^\dag (\textbf{r},
s_\alpha^+)= \sum_i q_i^\dag (s_\alpha^+) f_i(\textbf{r})$, it can
be shown that,
\begin{equation}
\label{eq: psim} \psi_i^{(m)} (s_\alpha) =  \sum_{j,k}
\Gamma_{ijk} \ q_j^\dag (s_\alpha^+) \ \psi_k^{(m-1)} (s_\alpha).
\end{equation}

In order to find $q_i^\dag (s)$ for the $(g-1)^{\rm{th}}$
generation, we first compute $q_i^\dag (s_{g-1}^+)$ by (\ref{eq:
qdi(s)}) and $\Lambda_i^\dag (s_g) = \delta_{i1}$.  Using
(\ref{eq: psim}) we can then iteratively compute $\psi_i^{(m)}
(s_{g-1})$ for all $m$ up to $\eta_{g-1}$.  Then we will have
$\Lambda_i^\dag (s_{g-1})$ and $q_i^\dag (s)$ for $s_{g-2} < s <
s_{g-1}$.  We can repeat this procedure until we have $q_i^\dag
(s)$ down to the first generation.  At this point we have computed
the probability of a chain diffusing from its branched, B-block
tips down to the end of A-block.  That is to say, we have computed
the single-chain partition function, $\mathcal{Q}/V =
q_1^\dag(s_0)$.

To find $q_i(s)$ we have solve the same matrix equation as
(\ref{eq: dqids}) except with a plus sign on the left-hand side.
Again, the transfer matrix for the ``reversed motion" A-block
solution is defined by,
\begin{equation}
T_{\rm{A},ij}(s'-s) \equiv \sum_k [O_A]_{ik} \exp \{ \mathcal{A}_k
(s' - s) \} [O_A]_{jk} \ ,
\end{equation}
and $T_{\rm{B},ij}(s'-s)$ is defined similarly.  The solution for
$q_i(s)$ is
\begin{equation}
\label{eq: qi(s)} q_i (s) = \left\{\begin{array}{ll} \sum_{j}
T_{\rm{A},ij} (s-s_1) \Lambda_j(s_1) \ , & \textrm{for $s_0 < s < s_1$} \ , \\ \\
 \sum_{j} T_{\rm{B},ij} (s-s_\alpha) \Lambda_j (s_\alpha)
 \ , & \textrm{for $s_{\alpha
- 1} < s < s_\alpha$} \ , \\ (\alpha \neq 1) \ ,
\end{array}\right.
\end{equation}
where $\Lambda_i (s_\alpha)$ are the branching point boundary
conditions,
\begin{equation}
\Lambda_i (s_{\alpha-1}) = V^{-1} \int d\textbf{r} \ q(\textbf{r},
s_{\alpha-1}^-) [q^\dag (\textbf{r}, s_{\alpha-1}^+)]^{\eta_\alpha
- 1} f_i(\textbf{r}).
\end{equation}
Similar to the identity (\ref{eq: psim}), it can be shown that
\begin{equation}
\label{eq: lami}
 \Lambda_i (s_{\alpha-1}) = \sum_{j,k}
\Gamma_{ijk} q_j (s_{\alpha-1}^-)  \psi_k^{(\eta_{\alpha}-1)}
(s_{\alpha-1}).
\end{equation}
We use the fact that $\Lambda_i (s_0) = \delta_{i1}$, the boundary
condition for the $s_0$ free end, to compute $q_i(s)$ from
(\ref{eq: qi(s)}) for the first generation.  We can then use
(\ref{eq: lami}) along with $q_i(s_1^-)$ and
$\psi_i^{(\eta_{1}-1)} (s_{1})$ to find $q_i(s)$ for the second
generation.  Repeating this process we can find $\Lambda_i
(s_{\alpha})$ and and $q_i(s)$ for all generations.

Using both $q_i(s)$ and $q_i^\dag (s)$ we can use (\ref{eq:
phiAreal}) and (\ref{eq: phiBreal}) to compute the Fourier
amplitudes of the monomer concentrations, $\phi_{A,i}$ and
$\phi_{B,i}$,
\begin{equation}
\phi_{A,i} = \frac{1}{q_1^\dag (s_0)} \int_{s_0}^{s_1} \! ds \
\sum_{j,k} q_j(s) q_k^\dag (s) \Gamma_{ijk} \ ,
\end{equation}
\begin{equation}
 \phi_{B,i} =  \frac{1}{q_1^\dag (s_0)}
\sum_{\alpha=2}^g \mathcal{N}_{\rm{B},\alpha}
\int_{s_{\alpha-1}}^{s_\alpha} \!\!\! ds \sum_{j,k} q_j(s)
q_k^\dag (s) \Gamma_{ijk} \ .
\end{equation}

Finally, the single-chain partition function, $\mathcal{Q}$, and
the Fourier densities are used to compute the mean-field free
energy per chain (up to an additive constant) are:
\begin{equation}
\frac{F}{n k_B T} = - \ln q_1^\dag (s_0) - \chi N \sum_i
\phi_{A,i} \phi_{B,i} \ .
\end{equation}
Now we need only to find the Fourier components of the field
configuration, $w_{A,i}$ and $w_{B,i}$, so that we satisfy the
self-consistency relations:
\begin{equation}
\label{eq: wis}
 w_{A,i}-w_{B,i} =\chi N(\phi_{B,i} - \phi_{A,i} )
\ ,
\end{equation}
\begin{equation}
\label{eq: phiis}
 \delta_{i1} = \phi_{A,i} +  \phi_{B,i}  \ .
\end{equation}
Clearly, we have $\phi_{A,1} = f$ and $\phi_{B,1} =  (1 - f) $, by
definition.  Moreover, we can set $w_{A,1} =\chi N (1 - f) $ and
$w_{B,1} =\chi N f $.  Since the monomer distributions,
$\phi_{A,i}$ and $\phi_{B,i}$, depend functionally on the fields,
$w_{A,i}$ and $w_{B,i}$, through $\mathcal{Q} [w_{A,i}, w_{B,i}]$,
Eqs. (\ref{eq: wis}) and (\ref{eq: phiis}) present a complicated
set of non-linear equations.  These are most easily solved by
computing $\phi_{A,i}$ and $\phi_{B,i}$, for some initial guess of
$w_{A,i}$ and $w_{B,i}$.  Then $w_{A,i}$ and $w_{B,i}$ can be
adjusted towards a solution of (\ref{eq: wis}) and (\ref{eq:
phiis}).  The computation then proceeds iteratively until the self-consistent solution is found.


\begin{thebibliography}{99}
\bibitem{FCC_note}
When considering close-packed lattices of spherical micelles, hexagonally close-packed and FCC lattices are known to have degenerate free-energies within mean field theory \cite{matsen_bates_macro_96_1}.  Here, we consider only the free-energy of the phase of $Fm\bar{3}m$ symmetry, the FCC phase.
\end{thebibliography}
\end{document}